\documentclass[a4paper,10pt,twoside]{just}

\usepackage{multicol}
\usepackage{graphicx}
\usepackage{booktabs}
\usepackage{amssymb,bm,mathrsfs,bbm,amscd}
\usepackage[tbtags]{amsmath}
\usepackage{lastpage}
\usepackage{CJK}

\def\beq{\begin{equation}}
\def\eeq{\end{equation}}
\def\bea{\begin{eqnarray}} 
\def\eea{\end{eqnarray}}

\def\gev{\rm GeV}

\def\mev{{\rm MeV}}

\def\eps{\varepsilon}

\begin{document}


\fancyhead[c]{\small  10th International Workshop on e+e- collisions from Phi to Psi (PHIPSI15)}
 \fancyfoot[C]{\small PHIPSI15-\thepage}
\title{\boldmath $\sin^2\theta_W$ theory and new physics}
\author{
Hye-Sung Lee\email{hlee@ibs.re.kr}
}

\maketitle

\address{
Center for Theoretical Physics of the Universe, IBS, Daejeon 34051, Korea
}

\begin{abstract}
After briefly discussing the importance of the precise measurement of the weak mixing angle, we discuss the implication of the dark $Z$ on the low-$Q^2$ parity tests.
The dark $Z$ is a very light (roughly, MeV - GeV scale) gauge boson, which couples to the electromagnetic current as well as the weak neutral current.
\end{abstract}

\begin{keyword}
parity test, dark force, dark photon, dark $Z$
\end{keyword}

\begin{pacs}
14.70.Pw, 11.30.Er
\end{pacs}

\begin{multicols}{2}

\section{Introduction}
In this article\footnote{This article and the presentation at PhiPsi 15 meeting (USTC, Hefei, China, September 23-26, 2015) are mainly based on the works and discussions with H. Davoudiasl and W. Marciano at Brookhaven National Lab, USA.}, we emphasize the importance of the low-$Q^2$ parity test for the new physics searches.
We illustrate our point with a specific example called the dark parity violation \cite{Davoudiasl:2012ag,Davoudiasl:2012qa,Davoudiasl:2014kua,Davoudiasl:2015bua}, which means the parity violation induced by a dark gauge boson.
This presentation shares some parts with Ref.~\cite{Lee:2014lga}, although updates and complementary descriptions are provided.

Let us briefly look back the history of the $\sin^2\theta_W$ physics.
It is well documented in the review~\cite{Kumar:2013yoa}, and we will go over only some part very briefly.
In 1961, Sheldon Glashow introduced the $SU(2)_L \times U(1)_L$ symmetry, which has a mixing between two neutral gauge bosons \cite{Glashow:1961tr}.
In 1967, Steven Weinberg added the Higgs mechanism with a Higgs doublet and a vacuum expectation value, establishing the mass relation $m_W = m_Z \cos\theta_W$ with the weak mixing angle $\theta_W$ \cite{Weinberg:1967tq}.
He also predicted the weak neutral current mediated by the $Z$ boson.
In 1973, the neutral current was discovered in the neutrino scattering experiments at the CERN Gargamelle detector \cite{Hasert:1973ff}.
Whether the $SU(2)_L \times U(1)_Y$ is a correct theory to describe this neutral current was not clear then though.
One of the features of the $SU(2)_L \times U(1)_Y$ was the mixing term in the weak neutral current interaction, proportional to $\sin^2\theta_W$, and the parity test measuring this $\sin^2\theta_W$ can possibly test the Standard Model (SM).

In 1978, SLAC E122 experiment using the polarized electron beam and the deuteron target measured the parity violation asymmetry, which gave $\sin^2\theta_W \approx 0.22 (2)$, agreeing to the SM \cite{Prescott:1978tm}.
It is noticeable that this establishment of the $SU(2)_L \times U(1)_Y$ by the SLAC parity test in 1978 occurred much earlier than the direct discovery of the $W$/$Z$ boson resonances at the CERN SPS experiments in 1983 \cite{Banner:1983jy,Arnison:1983mk}.
In 1979, after only one year of the SLAC parity test, Glashow, Salam, and Weinberg received the Nobel prize in physics.

The lessons we can learn from this history include
(i) the parity test (by the precise measurement of $\sin^2\theta_W$) can be a critical way to search for a new gauge interaction, and
(ii) its finding may precede the direct discovery of a gauge boson by the bump search.

Figure~\ref{fig:running} taken from Ref.~\cite{Davoudiasl:2015bua} shows the running of the $\sin^2\theta_W$ in the SM and the current experimental constraints.
While the current data are more or less consistent with the SM prediction with the given error bars, more precise measurements in the future experiments (red bars) may reveal potential new physics effects that were elusive for the current constraints.

\begin{center}
\includegraphics[width=0.45\textwidth]{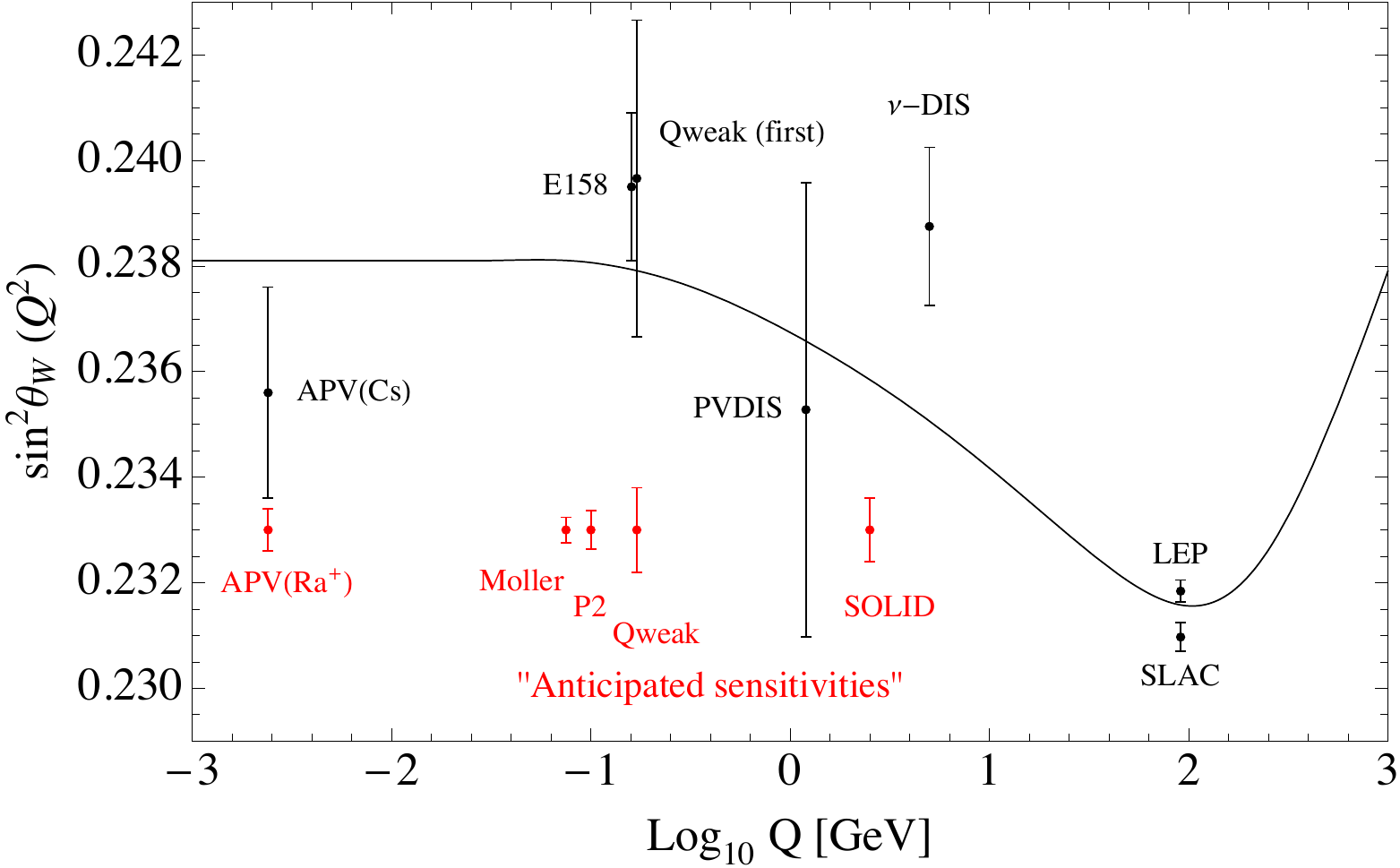}
\figcaption{\label{fig:running} The running of the $\sin^2\theta_W$ with the momentum transfer $Q$ in the SM and the current experimental constraints taken from Ref.~\cite{Davoudiasl:2015bua}. The red bars show the anticipated sensitivities in the future parity tests.}
\end{center}

\section{\boldmath Dark Photon vs. Dark $Z$}
The dark gauge boson (we use $Z'$ for its notation) is a hypothetical particle with a very small mass and a small coupling to the SM particles.
While the heavy $Z'$ (typically TeV scale) has been a traditional target of discovery (see Ref.~\cite{Langacker:2008yv} for a review), the light $Z'$ (typically MeV - GeV scale) is a recently highlighted subject with a growing interest (see Ref.~\cite{Essig:2013lka} for a review.)
For such a light particle to survive all experimental constraints, it should have extremely small couplings to the SM particles.

There are number of dark force models in the literature, but we consider only two of them.
Both models commonly assume the kinetic mixing between the $U(1)_Y$ and the $U(1)_{\rm dark}$ \cite{Holdom:1985ag}.
The SM particles have zero charges under the new gauge group $U(1)_{\rm dark}$, yet the gauge boson $Z'$ of the $U(1)_{\rm dark}$ can still couple to the SM fermions through the mixing with the SM gauge bosons.

One model is the dark photon \cite{ArkaniHamed:2008qn}, which couples only to the electromagnetic current at the leading order.
Another is a relatively new model, the dark $Z$ \cite{Davoudiasl:2012ag}, which couples to the electromagnetic current as well as the weak neutral current.
Their interactions are given by
\begin{eqnarray}
{\cal L}_{{\rm dark}~\gamma} &=& - \eps e J^\mu_{EM} Z'_\mu \label{eq:darkPhoton} \\
{\cal L}_{{\rm dark}~Z} &=& - \left[ \eps e J^\mu_{EM} + \eps_Z (g / 2 \cos\theta_W) J^\mu_{NC} \right] Z'_\mu \label{eq:darkZ}
\end{eqnarray}
with $J_\mu^{EM} = Q_f \bar f \gamma_\mu f$ and $J_\mu^{NC} = (T_{3f} - 2 Q_f \sin^2\theta_W ) \bar f \gamma_\mu f - (T_{3f}) \bar f \gamma_\mu \gamma_5 f$.
$\eps$ and $\eps_Z$ are the parametrization of the effective $\gamma - Z'$ mixing and $Z - Z'$ mixing, respectively.

The difference of the two models comes from how the $Z'$ gets a mass or the details of the Higgs sector.
Because of the $Z$ coupling, the $Z'$ in the dark $Z$ model inherits some properties of the $Z$ boson such as the parity violating nature.
In a rough sense, the dark photon is a heavier version of the photon, and the dark $Z$ is a lighter version of the $Z$ boson.

Because of the new coupling, some experiments that are irrelevant to the dark photon searches are relevant to the dark $Z$ searches \cite{Davoudiasl:2012ag,Davoudiasl:2012qa,Davoudiasl:2014kua,Davoudiasl:2015bua,Kong:2014jwa,Davoudiasl:2013aya,Davoudiasl:2014mqa,Kim:2014ana}.
They include the low-$Q^2$ parity test, which will be discussed later in this article.

\section{\boldmath Bump Hunt}
There are many ongoing and proposed searches for the dark force in the labs around the world \cite{Essig:2013lka}.
A particularly attractive feature about the dark force is that it is one of the rare new physics scenarios that can be tested/discovered at the low-energy experiments, which are typically built for nuclear or hadronic physics.
Of course, it is possible because the dark force carrier $Z'$ is very light (MeV - GeV scale).

Figure 2 in Ref.~\cite{Soffer:2015kpa} shows the parameter space of the dark photon with the current bounds.
The bounds come from the electron \cite{Davoudiasl:2012ig,Endo:2012hp} and muon \cite{Gninenko:2001hx,Fayet:2007ua,Pospelov:2008zw} anomalous magnetic moments, fixed target experiments \cite{Abrahamyan:2011gv,Merkel:2014avp}, beam dump experiments \cite{Andreas:2012mt}, meson decays \cite{Bjorken:2009mm,Babusci:2012cr,Agakishiev:2013fwl,Adare:2014mgk,ALICE,Batley:2015lha}, $e^+e^-$ collision ($e^+e^- \to \gamma + \ell^+\ell^-$) experiments \cite{Babusci:2014sta,Lees:2014xha}.
Except for the anomalous magnetic moments, the searches are all based on the dilepton searches from the $Z'$, that is the bump hunt.

If we put some of these experimental efforts on the map (Figure~\ref{fig:global}), we can see the search is practically a global activity.
Quite obviously, we are going through a very exciting time with so many contemporary searches to find a new fundamental force of nature.

\begin{center}
\includegraphics[width=0.45\textwidth]{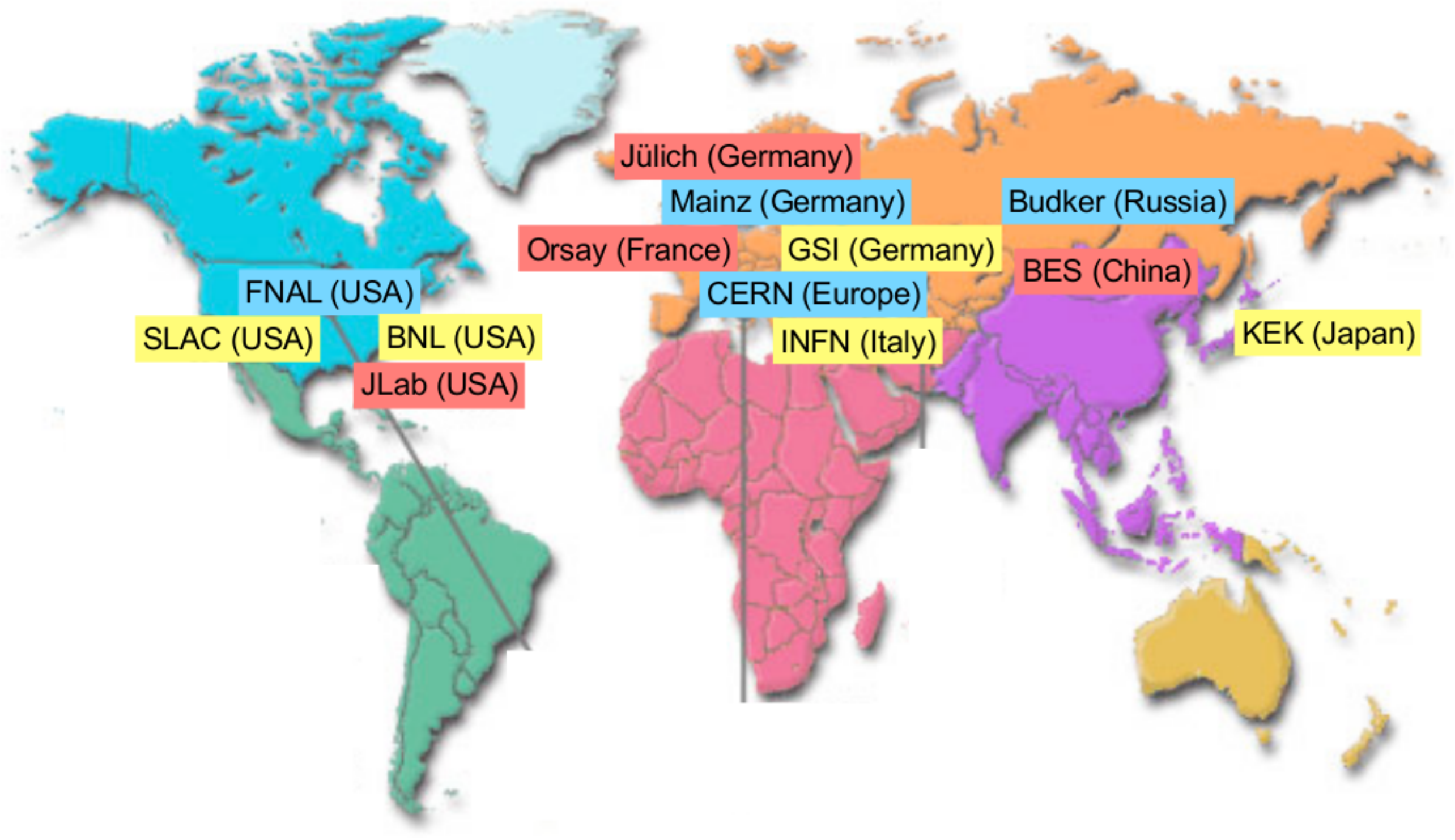}
\figcaption{\label{fig:global} Dark force searches all around the world.}
\end{center}

\section{\boldmath Low-Energy Parity Test}
Now, we discuss the low-energy parity test \cite{Davoudiasl:2012ag,Davoudiasl:2012qa,Davoudiasl:2014kua,Davoudiasl:2015bua} as another mean to search for the dark force.
 The presence of the dark $Z$ modifies the effective lagrangian of the weak neutral current scattering,
\bea
{\cal L}_{\rm eff} &=& - \frac{4 G_F}{\sqrt{2}} J^\mu_{NC} ( \sin^2\theta_W ) J_\mu^{NC} ( \sin^2\theta_W ) \\
G_F \!\!\!\! &\to& \!\!\!\! \left(1 + \delta^2 \frac{1}{1 + Q^2 / m_{Z'}^2} \right) G_F \\
\sin^2\theta_W \!\!\!\! &\to& \!\!\!\! \left(1 - \eps\delta \frac{m_Z}{m_{Z'}} \frac{\cos\theta_W}{\sin\theta_W} \frac{1}{1 + Q^2 / m_{Z'}^2} \right) \sin^2\theta_W ~~
\eea
where $Q$ is the momentum transfer between the two neutral currents, and $\delta$ is a reparametrization of the $\eps_Z$ with $\eps_Z \equiv (m_{Z'} / m_Z) \delta$.
One salient feature is that these shifts are sensitive only to the low-$Q^2$ (low momentum transfer).
Thus, the dark $Z$ effectively changes the weak neutral current scattering, including the effective $\sin^2\theta_W$, which describes the parity violation, but only for the low momentum transfer.

Figure~\ref{fig:darkRunning1} (from Ref.~\cite{Davoudiasl:2014kua} with a slight modification) shows an example of how the effective $\sin^2\theta_W$ changes with $Q$ in the presence of a dark $Z$ for $m_{Z'} = 100~\mev$ (blue band), $200 ~\mev$ (red band) cases.
Although there are some details in the figure, the important point is that the deviations appear only in the low $Q$ values, roughly $Q \le m_{Z'}$.
They never appears in the high $Q$ values relevant to the high-energy experiments, which tells us that we need low-energy experiments to see the dark $Z$ mediated scattering effects.

\begin{center}
\includegraphics[width=0.45\textwidth]{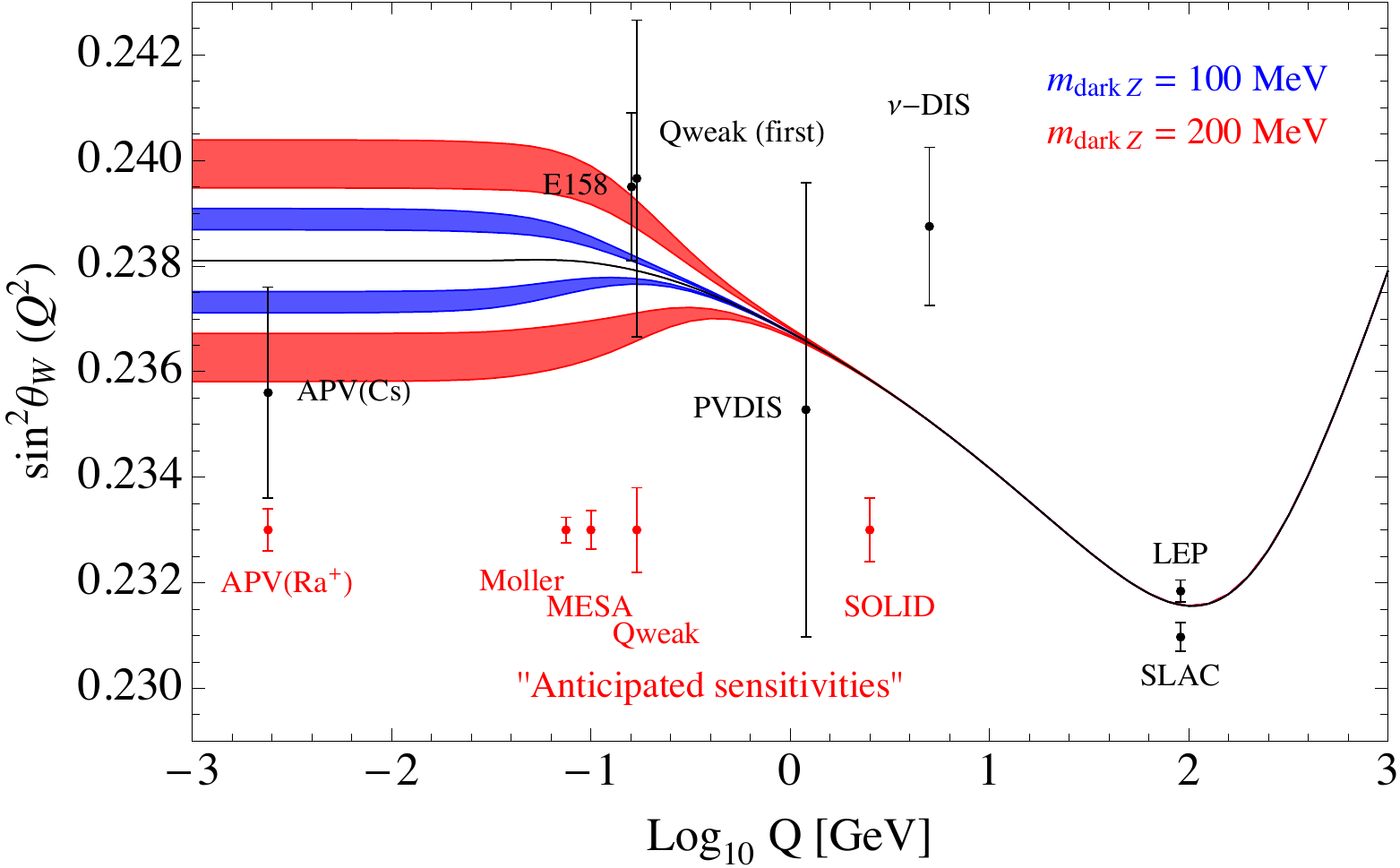}
\figcaption{\label{fig:darkRunning1} Effective $\sin^2\theta_W$ running taken from Ref.~\cite{Davoudiasl:2014kua}. Dark $Z$ of 100 MeV (blue) and 200 MeV (red) were taken. Note that the deviations appear only in the low-$Q^2$ regime ($Q^2 \le m_{Z'}^2$).}
\end{center}

In this region, non-perturbative QCD contributions to the SM value become important. They have traditionally been determined utilizing dispersion relations \cite{Erler:2004in,Jegerlehner:2011mw}.
Recently, also first-principle lattice QCD determinations of the leading-order hadronic effects have become available \cite{Burger:2015lqa,Francis:2015grz}.

For the low-$Q^2$ parity tests, one can use the atomic parity violation in {\rm Cs} \cite{Gilbert:1986ki,Wood:1997zq,Bennett:1999pd}, {\rm Ra$^+$} ion \cite {NunezPortela2014,Jungmann:2014kia} or the low-$Q^2$ polarized electron scattering experiments SLAC E158 \cite{Anthony:2005pm}, JLAB Qweak \cite{Androic:2013rhu}, JLAB Moller \cite{Mammei:2012ph} and Mainz P2 \cite{MESA}.
The possible deviations due to the dark $Z$ can be large enough to be observed with the future experiments.

For the intermediate scale $Z'$ of $m_{Z'} \approx {\cal O}(10) ~\gev$, the deep inelastic scattering experiments such as JLAB PVDIS \cite{Wang:2014bba} and JLAB SOLID \cite{Reimer:2012uj} may be also sensitive.
In fact, as Fig.~\ref{fig:darkRunning2} taken from Ref.~\cite{Davoudiasl:2015bua} shows, the intermediate scale $Z'$ can address the NuTeV ($\left< Q \right> \approx 5 ~\gev$) anomaly \cite{Zeller:2001hh}.

\begin{center}
\includegraphics[width=0.45\textwidth]{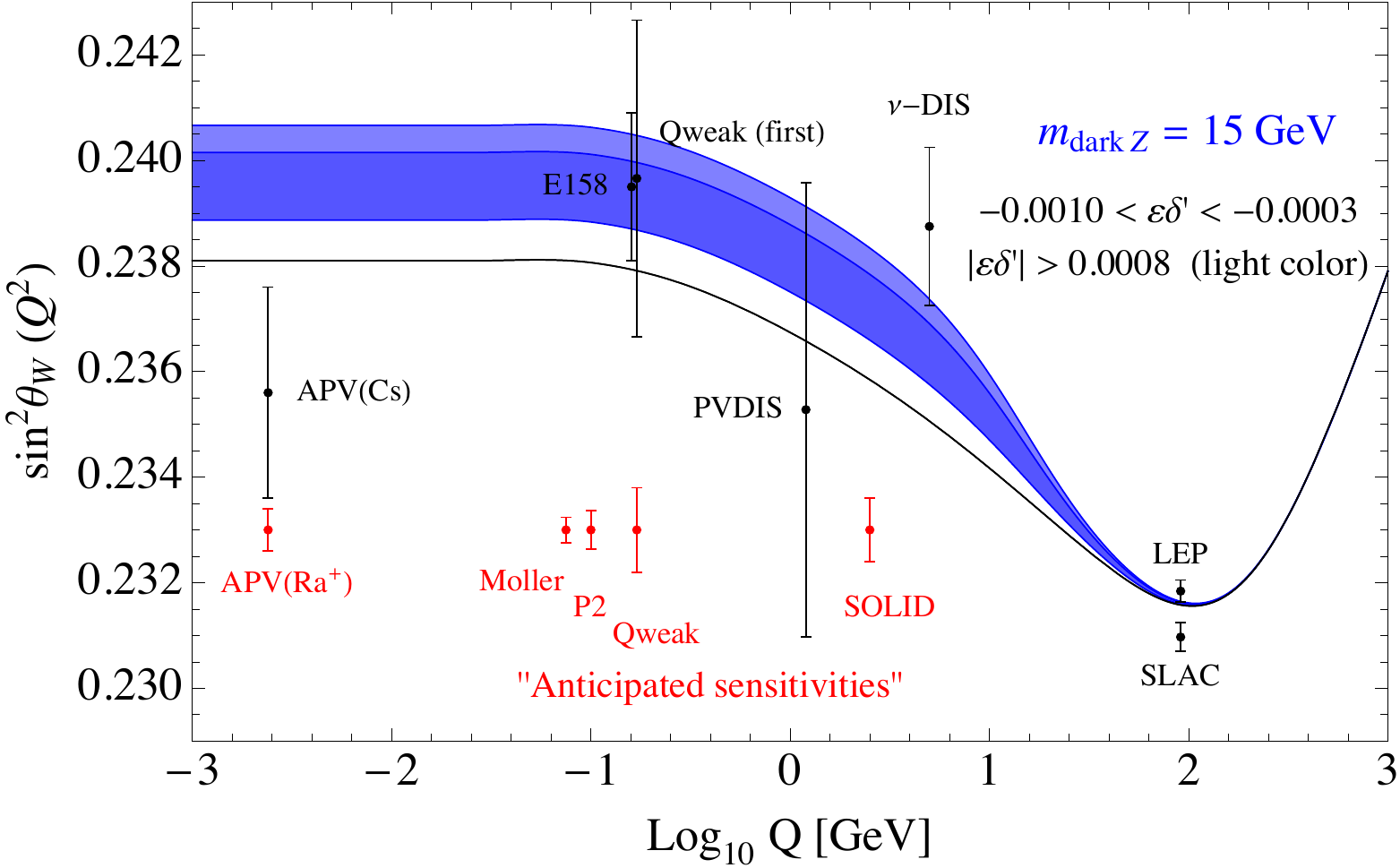}
\figcaption{\label{fig:darkRunning2} The 15 GeV dark $Z$ case taken from Ref.~\cite{Davoudiasl:2015bua}. The NuTeV anomaly can be addressed by this intermediate scale dark $Z$.}
\end{center}

\section{\boldmath Summary}
The parity test by precise measurement of the $\sin^2\theta_W$ has been important in studying new gauge interactions.
Especially, it critically helped establishing the $SU(2)_L \times U(1)_Y$ electroweak theory.
There is a growing interest in the dark gauge interaction (mediated by a light $Z'$ gauge boson) around the world partly because many existing low-energy facilities can join the searches.
While most searches of the light $Z'$ are based on the direct bump searches, the parity tests in the low-$Q^2$ (such as the atomic parity violation, polarized electron scattering, deep inelastic scattering) are important and complementary searches for the dark force.
The latter are also independent of the $Z'$ decay branching ratios.

If the history may repeat, the dark force evidence from the low-$Q^2$ parity test might precede the discovery of a new resonance just like what happened in the electroweak interaction case.

\vspace{2mm}
\acknowledgments{I am very grateful to H. Davoudiasl and W. Marciano for a long-term collaboration. This work was supported in part by the IBS (Project Code IBS-R018-D1).}

\end{multicols}

\vspace{0mm}

\begin{multicols}{2}

\end{multicols}

\clearpage

\end{document}